\documentclass[12pt,preprint]{aastex}

\lefthead{Bekki,  Chiba, \& McClure-Griffiths}
\righthead{The Magellanic impact}

\begin{document}
\title{The Magellanic impact: Collision between the outer
Galactic H~{\sc i}  disk and the leading arms of the Magellanic stream}

\author{Kenji Bekki\altaffilmark{1}} 

\author{Masashi Chiba\altaffilmark{2}} 

\and

\author{N. M. McClure-Griffiths\altaffilmark{3}} 

\altaffiltext{1}{School of Physics, University of New South Wales,
Sydney 2052, Australia, bekki@phys.unsw.edu.au}
\altaffiltext{2}{Astronomical Institute,
Tohoku University, Sendai, 980-8578, Japan, chiba@astr.tohoku.ac.jp}
\altaffiltext{3}{Australia Telescope National Facility, CSIRO, P.O. Box 76,
Epping NSW 1710, Australia, Naomi.McClure-Griffiths@csiro.au}

\begin{abstract}

We  show that collisions between the outer Galactic H~{\sc i}  disk
and the leading arms (LAs) of the Magellanic stream (MS) can 
create giant H~{\sc i}  holes and chimney-like structures in the disk.
Based on 
the results of our  N-body simulations on the last 2.5 Gyr evolution
of the Large and Small Magellanic Clouds (LMC and SMC, respectively)
interacting with the Galaxy,
we  investigate  when and where   
the LAs can pass through 
the Galactic plane after the MS formation. 
We then investigate hydrodynamical interaction between  LAs
and the Galactic H~{\sc i}  disk (``the Magellanic impact'')
by using our new hydrodynamical simulations
with somewhat idealized models of the LAs.
We find that about 1-3\% of the initial gas mass of the SMC,
which consists of the LAs,
can pass through the outer part ($R=20-35$ kpc)
of the Galactic H~{\sc i}  disk  about 0.2 Gyr ago.
We also find that the Magellanic impact can push out
some fraction ($\sim 1$\%)  of the outer Galactic H~{\sc i}   disk to form
1-10 kpc-scale H~{\sc i}  holes and chimney-like bridges between
the LAs and the disk. 

\end{abstract}


\section{Introduction}

H~{\sc i}  gas in the Magellanic stream, its leading arms,
and the Magellanic bridge have long been discussed 
in the context of dynamical and hydrodynamical interaction
between the Magellanic Clouds (MCs) and the Galaxy
(e.g., Murai \& Fujimoto 1980; Putman et al. 1998;
Mastropietro et al. 2005; Muller \& Bekki 2007).
Such intense interactions between the three galaxies
have been also  suggested to be closely associated 
not only with
the long-term star formation histories
of the LMC and the SMC (e.g.,  Harris \& Zaritsky 2004;
Cioni et al. 2006)
but also with structural and kinematical properties
of the MCs (e.g., Bekki \& Chiba 2005, BC05).
It is however not fully understood how
the interactions can change 
dynamical  and chemical properties  and star formation
histories of  the MCs
(e.g., Westerland 1997).

Although many previous numerical/theoretical works
discussed influences of the Galaxy on the evolution
of the MC (e.g., Gardiner \& Noguchi 1995;
BC05; 
R\. u\v zi\v cka et al. 2007),
only a few works investigated the possible
gravitational influences of the MCs on the 
dynamical evolution of  the Galactic {\it the stellar disk} 
(e.g., Weinberg 1998; Tsuchiya 2002)
and the gaseous one (Weinberg \& Blitz 2006).
For example, these works discussed whether
the observed Galactic warp can be due to the dynamical
interaction between the LMC and the Galaxy (e.g., Tsuchiya 2002).
Although previous numerical and observational studies
on the MS formation
suggested possible interaction between the LAs and the Galaxy
(e.g., Yoshizawa \& Noguchi 2003; Br\"uns et al. 2005;
Connors et al. 2006),
the details of the interaction processes have not been
investigated yet.
In particular, 
it is totally unclear {\it how the gaseous components of the
Galaxy are influenced by the gaseous components of the MCs}.

Recent observational studies on the physical properties
of the outer H~{\sc i}  disk of the Galaxy have 
reported a  significant ($\sim 10$ km s$^{-1}$)
velocity perturbation in the outer H~{\sc i}  spiral
arm of the Galaxy at $l=275^{\circ}-295^{\circ}$ 
(McClure-Griffiths et al. 2004, M04).
Furthermore,
it remains unclear why the outer H~{\sc i}  disk ($19<R<28$ kpc) 
appears to show  significantly lower densities 
between the Galactic longitudes $l=285^{\circ}$ and $300^{\circ}$
(See Figure 2 in M04). 
No theoretical works have however discussed so far the origin of
these unique properties in M04.

The purpose of this Letter is to show, for the first time,
how collision between the LAs and the Galactic H~{\sc i}  disk
(hereafter referred to as ``the Magellanic impact'')
influences the evolution of the Galactic H~{\sc i}  disk.
We first show when and how the LAs can pass through the
Galactic disk based on our collisionless  N-body simulations
on the evolution of the MCs for the last 2.5 Gyr 
(e.g., Bekki \& Chiba 2007, BC07).
We then show that the Magellanic impact
can create unique, kpc-scale H~{\sc i}  holes and chimney-like
bridges connecting the LAs and the outer part of the
Galactic H~{\sc i}  disk
based on new yet 
somewhat idealized hydrodynamical simulations on the
Magellanic impact.

\section{The  two-fold model}
The present investigation is two-fold as follows.
We first investigate when and where the LAs can pass through the
Galaxy during the LMC-SMC-Galaxy interaction based on purely
collisionless N-body simulations similar to our previous ones
(BC07).  
In this first investigation,
we focus exclusively on the locations and the masses
of the LAs at the epochs of LAs' passage through the Galactic disk,
because previous models did not clearly describe these.
We then investigate how the Magellanic impact influences
the outer part of the Galactic H~{\sc i}  disk for a reasonable
set of initial conditions of the Magellanic impact derived by
the above first investigation by using our new hydrodynamical
simulations. Since numerical methods and techniques for the
first collisionless N-body
simulations and the second GRAPE-SPH ones  are already given in
BC07 and Bekki \& Chiba (2006), respectively, 
we here briefly describe them.

\subsection{The LA formation}

We first determine the most
plausible and realistic orbits of the MCs
by using `` the backward integration
scheme'' (for orbital evolution  of the MCs) by Murai \&  Fujimoto (1980)
for the last 2.5 Gyr and then perform the first set
of collisionless N-body simulations on  the
evolution of the MCs using GRAPE
systems (Sugimoto et al.1990).
The total masses of the LMC  and the SMC
are set to be $2.0 \times 10^{10} {\rm M}_{\odot}$
and $3.0 \times 10^{9} {\rm M}_{\odot}$,
respectively,  in all models.
The SMC is represented by a fully self-consistent dynamical
model composed of a  dark matter halo, a  stellar disk or spheroid,
and a  gas disk
whereas 
the LMC is represented by a point mass.

The mass fraction of the baryonic component (i.e., stars and gas)
and the gas mass fraction of the baryon
in the SMC
are considered to be free parameters and represented by $f_{\rm b}$
and $f_{\rm g}$, respectively.
The ratio ($s$) of the gas disk size ($R_{\rm g, SMC}$) to
the stellar one ($R_{\rm s, SMC}$) in the SMC is also
considered to be a free parameter that can effect the
morphologies of the MS and the LAs.
Although we have investigated models with different 
$f_{\rm b}$, $f_{\rm g}$, $R_{\rm g, SMC}$, and $s$,
we focus on  the results of a LA model
with $f_{\rm b}=0.33$, $f_{\rm g}=0.20$, $R_{\rm g,SMC}=7.5$ kpc,
and $s=4.0$: Results of the other models will be given
in our future papers (Bekki \& Chiba in preparation).
This model is simply referred to as the LA model.

We use the same coordinate system  $(X,Y,Z)$
(in units of kpc) as those used
in BC05 and BC07.
The adopted current positions
are $(-1.0,-40.8,-26.8)$  for the LMC
and $(13.6,-34.3,-39.8)$
 for the SMC and
the adopted
current  Galactocentric radial velocity of the LMC (SMC)
is 80 (7) km s$^{-1}$.
Current velocities of the LMC
and the SMC in the
Galactic ($U$, $V$, $W$) coordinate
are assumed to be
(-5,-225,194) and  (40,-185,171) in units of km s$^{-1}$,
respectively. 
As shown  by previous studies (e.g., Gardiner \& Noguchi 1995),
the MS and the LAs can be formed
for the adopted orbits of the MC.
We do not intend to investigate the models with
initial velocities consistent with the latest
proper motion data (e.g., Kallivayalil et al. 2006).
We investigate the accretion rate 
($\dot{m}_{\rm SMC,g}$)
of the SMC's  gas particles
initially in the LA  onto the outer part of the
Galaxy,
the total masses of the LAs
passing though the Galactic disk  ($m_{\rm LA}$),
the epochs of the passage  ($T_{\rm LA}$), and the inclination angles
of the LAs with respect to the Galactic disk 
at $T=T_{\rm LA}$ (${\theta}_{\rm LA}$).

\subsection{The tube model for the Magellanic impact}

We investigate hydrodynamical interaction between the LAs and the outer
part of the Galactic H~{\sc i}  disk for models with different 
initial properties of the LAs
(e.g., $m_{\rm LA}$ and ${\theta}_{\rm LA}$)
derived by the first set of collisionless simulations of the LA formation. 
In this second set of GRAPE-SPH simulations,
the gaseous stream of the  LA is assumed to be represented by a long 
``tube'' with a size of $r_{\rm LA}$,  a length of $l_{\rm LA}$,
a mass of $m_{\rm LA}$,
a position (with respect to the Galactic center) 
of ${\bf   x}_{\rm LA}$,
an inclination angle of ${\theta}_{\rm LA}$,
and a velocity of $v_{\rm LA}$ ($\ge 0$).
We adopt the ring model (BC06) for 
the outer part  
(20 kpc $\le R \le $ 40 kpc) of the Galactic H~{\sc i}  disk  with 
a uniform
distribution so that we can focus solely on
the evolution of {\it the outer part} of the Galactic H~{\sc i}  disk
during the Magellanic impact. Considering the observational
fact that the total H~{\sc i}  mass of the Galaxy is about 
$4 \times 10^9 {\rm M}_{\odot}$ 
(e.g., van der Kruit 1989),
the total gas mass of the ring for 20 kpc $\le R \le $ 40 kpc
is assumed to be $1.4 \times 10^9 {\rm M}_{\odot}$.
We assume that the size of the stellar disk
of the Galaxy ($R_{\rm s,MW}$) is 17.5 kpc 
and that the Galaxy is  composed only of gas
outside $R_{\rm s,MW}$.

We here discuss 
the results of two representative models, Model A and B,
among those that we have investigated.
Firstly we show the result of the Model A
with $r_{\rm LA}=2.1$ kpc, $l_{\rm LA}=21$ kpc, 
$m_{\rm LA}=10^7 {\rm M}_{\odot}$,
${\bf   x}_{\rm LA}$=(31.5 kpc, 0 kpc, -10.5 kpc),
${\theta}_{\rm LA}=60^{\circ}$, and 
$v_{\rm LA}=393$ km s$^{-1}$ in \S 3.
We chose the above ${\bf   x}_{\rm LA}$ so that
we can show more clearly the evolution
of the outer part of the Galactic disk
projected onto the $X$-$Y$ and $X$-$Z$ planes: the adopted values are 
slightly different from  the predicted locations of the LA.
We then discuss the results of the Model B 
with $r_{\rm LA}=2.1$ kpc, $l_{\rm LA}=63$ kpc,
$m_{\rm LA}=10^8 {\rm M}_{\odot}$,
${\bf   x}_{\rm LA}$=(21.0 kpc, -15.8 kpc, -10.5 kpc),
${\theta}_{\rm LA}=60^{\circ}$, and
$v_{\rm LA}=314$ km s$^{-1}$ in \S 4. 
This model can reproduce best the  observations by M04.

\section{Results}

Figure 1 shows that the LAs composed mainly of two gaseous
streams can pass through 
the Galactic plane about 0.2 Gyr (i.e., $T\sim-0.2$ Gyr)
in the  LA model.
The high-density tip of the LA passes through the outer part
($R>20$ kpc) of the Galaxy with a high ${\theta}_{\rm LA}$
(the inclination angle between the LA and the Galactic plane)
and a high vertical velocity of $v_{\rm LA, z} = 240$  km s$^{-1}$.
About 1.4 \% of the inital  gas particles of the SMC can pass though
the outer region ($R_{\rm s, MW} \le R \le 2R_{\rm s, MW}$)
for the last $\sim 0.3$ Gyr and 
the vertical velocities ($v_{\rm LA, z}$)  of the gas particles 
in the LAs are rather high and
range from 186 km s$^{-1}$ to 317 km s$^{-1}$.
The accretion rate ${\dot m}_{\rm SMC,g}$ reaches  a peak
at $T \sim -0.2$ Gyr 
with the peak value of $\sim 0.4 {\rm M}_{\odot}$  yr$^{-1}$.
It is found that the projected distribution
of  particles passing through the Galactic disk at $T \sim -0.2$ Gyr
are  well confined in the sense that most particles
are located at $-10$ kpc $\le X \le$ $-5$ kpc and
at $-20$ kpc $\le Y \le$ $-15$ kpc.

Figure 2 shows how the Magellanic impact changes the vertical
structure of the outer part of the Galactic gas disk 
in the Model A. 
Although the Magellanic impact can not change the global structure
of the Galactic gas disk (e.g.,  warps
and spiral arms) owing to the adopted small mass
of the LA ($m_{\rm LA}=10^7 {\rm M}_{\odot}$), 
it can push out a small fraction $\sim 1$\%) of the gas disk
and  form chimney-like
bridges connecting the LA and the Galaxy.
The bridges can be seen  above the Galactic disk (i.e., $Z>0$)
not only in this Model A  but also in  other models,
which is thus a robust prediction that can be tested against
ongoing and future H~{\sc i}  observations (e.g., M04
and McClure-Griffiths et al. 2006, M06). 
The gaseous particles in the bridges for 2 kpc $\le Z \le$ 10 kpc 
at $T=164$ Myr have large (positive) vertical velocities
($95$ km s$^{-1}$ on average)
that are significantly different from those of the particles
within the Galactic H~{\sc i}  disk.

Figure 3 shows that as a result of the Magellanic impact,
a giant H~{\sc i}  hole with a size of $\sim 10$ kpc  
can be formed in the outer part ($R \sim 35$ kpc) of the
Galactic H~{\sc i}  disk.
The H~{\sc i}  hole can  be neither destroyed nor significantly elongated 
owing to  the shear motion of the Galactic gas disk within 
a time scale of 0.2 Gyr (i.e., the time interval between
the Magellanic impact and the present)
so that it can be clearly seen in the final snapshot of the model.
Only the right part of the hole 
(i.e., $X \sim 30$ kpc and $Y \sim 22$ kpc)
clearly shows a ridge with a significantly
higher gas density (${\mu}_{\rm g} \sim 0.8 {\rm M}_{\odot}$ pc$^{-2}$).
Such giant H~{\sc i}  holes   in the very {\it outer} part of 
the Galaxy are highly unlikely to form via feedback effects
of massive stars and supernovae, 
because the rate of massive  
star formation at large radii is very low in such low density regions.

The surface gas densities
along azimuthal angles (${\mu}_{\rm g} (\theta)$)
averaged over a range of 
30 kpc $\le$ $R$ $\le$  40 kpc
in the Galactic disk   
are found to range from $0.16 {\rm M}_{\odot}$ pc$^{-2}$
to $0.48 {\rm M}_{\odot}$ pc$^{-2}$ 
with the mean of ${\mu}_{\rm g} (\theta)$ 
being $0.40 {\rm M}_{\odot}$ pc$^{-2}$  in this model.
The minimum value is found where the giant H~{\sc i} hole
exists, which suggests that the line-of-sight column
density 
(with respect to the Sun)
in the direction to an  H~{\sc i}  hole 
can be a factor of $2-3$ smaller than those 
in other directions.
We discuss these results in terms of observations by M04 
later in \S 4.
We confirm that 
the models with
$m_{\rm LA}=10^5-10^6 {\rm M}_{\odot}$ do not show any kpc-scale
holes with remarkable chimney-like structures
owing to the rather weak dynamical impacts of the LA
on the Galactic H~{\sc i}  disk.

\section{Discussion and conclusions}

We first have shown that the Magellanic impact can create
10kpc-scale H~{\sc i}  holes and bridges connecting 
the LA and the Galaxy.
 The physical properties of the simulated unique structures
in the outer part ($R\sim 30$ kpc) of the Galactic gas disk 
can be tested against previous and ongoing observational studies on
H~{\sc i}  properties of the Galaxy 
(e.g., M04; Levine et al. 2006; M06).
Although  
detailed comparison of ongoing and future observations will be given  
in our future papers, here we briefly discuss previous observations  
shown in Figures 1 and 2 of M04 compared to 
the  Model B 
in a more quantitative manner.

Figure 4 clearly shows the longitude-velocity  ($l-v$)
 diagram of particles
of the Galactic gas disk in the Model B, where
$v$ denotes the line-of-sight velocities
of the particles  with respect to the Sun.
Clearly the  $l-v$ diagram shows a bifurcation
around at $l=280^{\circ}-300^{\circ}$,
the most clearly  at $l \sim 290^{\circ}$.
This bifurcation is due to the dynamically disturbed gas disk 
of the Galaxy 
by the Magellanic impact and  may correspond to 
a part of the observed  significant ($\sim 10$ km s$^{-1}$)
velocity perturbation in the outer H~{\sc i}  spiral
arm of the Galaxy at $l=275^{\circ}-295^{\circ}$ 
(see Figure 1 of M04).
Figure 4 also shows a significantly less populated 
region  at  $l=280^{\circ}-300^{\circ}$
in  the particle distribution on the $(l-b$) plane
owing to the
presence of the giant H~{\sc i}  hole there created by
the Magellanic impact.
This is broadly consistent with observations shown
in Figure 2 of M04, which shows that
the outer H~{\sc i}  disk of the Galaxy 
shows  significantly lower densities 
between the Galactic longitudes $l=285^{\circ}$ and $300^{\circ}$.

Velocities of particles pushed out from the Galactic plane
can be more than 50 km s$^{-1}$ higher than those within the
Galactic disk in Figure 4, which can not be seen in Figure 1 of M04.
This apparent inconsistency possibly means that
such high-velocity
gas with low-density and low-mass ($\sim 10^3 {\rm M}_{\odot}$)
can be soon changed into ionized gas owing to possible interaction
with warm/hot  halo gas of the Galaxy so that it can not be
observed as H~{\sc i}.
Although the observed $l-v$ diagram  in M04 does not clearly show bifurcation,
the H~{\sc i} seems to be distributed on two belts for 
$260^{\circ} <  l < 295^{\circ}$ and $v>$  60 km s$^{-1}$.
It remains observationally unclear whether this intriguing distribution
can correspond to a bifurcated distribution of H~{\sc i} on the $l-v$ diagram.

The direction
of the orbit of the LA (with respect to the Galaxy)
can significantly change at the Galactic plane (i.e., $b=0^{\circ}$)
as a result of hydrodynamical interaction between the LA and
the Galactic H~{\sc i}  disk.  The orbital change at  $b=0^{\circ}$
due to the back reaction of the Magellanic impact 
can explain why the observed distribution
of the LA in the Galactic coordinate shows a  ``kink'' 
at  $b=0^{\circ}$
(e.g., Br\"uns et al. 2005).
Furthermore  the total H~{\sc i}  mass of the LA can be significantly
reduced by photoionization of the H~{\sc i}  gas of the LA by
the warm gas of the Galactic disk
during the Magellanic impact.
This significant reduction of the H~{\sc i}  mass in the LA
can solve the well known problem as to the observed small mass
ratio ($\sim 0.2$) of the LA to the MS (e.g., Br\"uns et al. 2005).

It is possible  that  the Magellanic impact
can trigger some amount  of  star formation in the outer part
of the  Galactic gas disk
owing to collisions between gas clouds from the Galaxy and
those from the LA.  
We  thus suggest that the presence (or the absence)
of B-type stars with ages of $0.1-0.2$ Gyr at $270^{\circ} <  l < 300^{\circ}$
and $b \sim 0^{\circ}$  would prove (or disprove) the past
star formation activities of the Galaxy induced by the Magellanic impact.
We also suggest that
hydrodynamical interactions between gaseous  tidal tails and debris  from
disrupting low-mass galaxies (like the SMC)
and outer parts of galactic H~{\sc i}  disks
can be proved by observations on detailed spatial distributions  of  H~{\sc i}
in galaxies.
In particular,  future structural and kinematical studies on
the observed giant H~{\sc i} holes in the very outer parts of galaxies
(e.g., NGC 6822;  de Blok \& Walater 2003) 
will allow us to discuss whether or not hydrodynamical interaction
between gaseous tidal tails and outer gas disks of galaxies can be important
for the evolution of the outer disks.

\acknowledgments
We are  grateful to the anonymous referee for valuable comments,
which contribute to improve the present paper.
K.B. acknowledges the financial support of the Australian Research
Council throughout the course of this work.
The numerical simulations reported here were carried out on GRAPE
systems kindly made available by the Center for Computational
Astrophysics (CfCA)
at National Astronomical Observatory of Japan (NAOJ).

\clearpage

\begin{figure}
\epsscale{.50}
\plotone{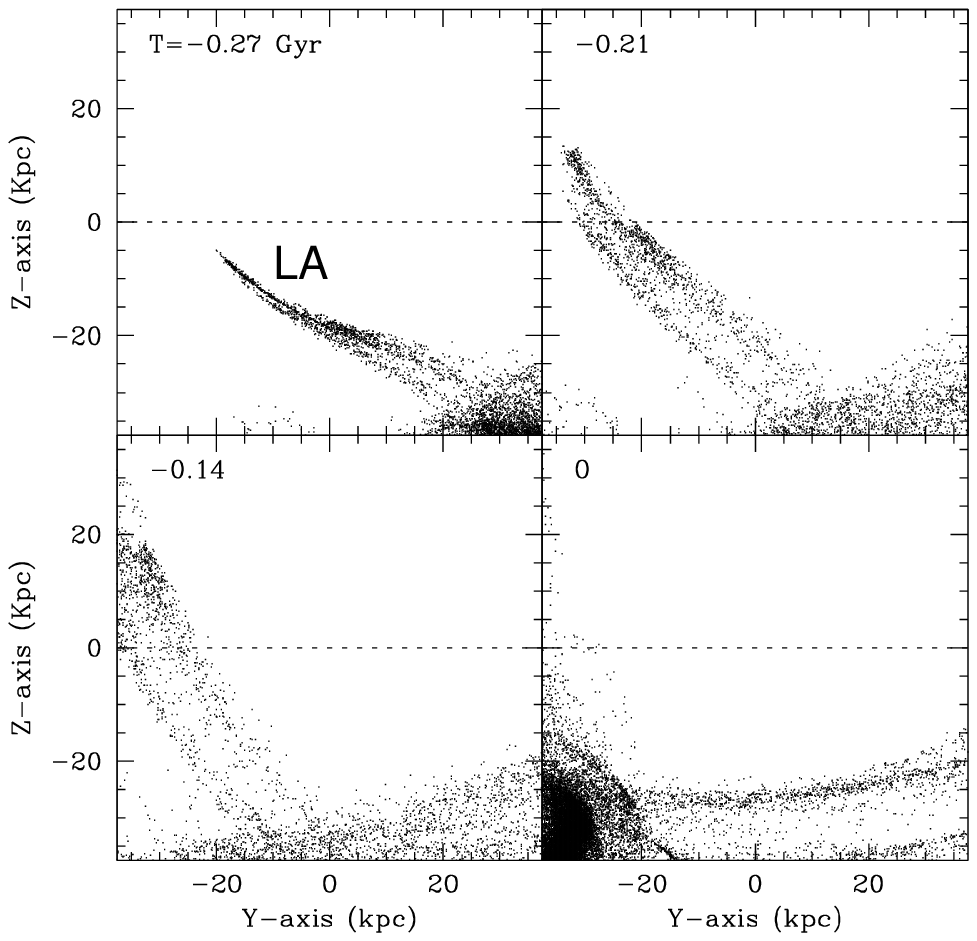} 
\caption{
Time evolution of the LAs of the MS 
projected onto the $Y$-$Z$ plane 
for the last 0.27 Gyr
in the   LA model. 
$T=0$ means the present while the negative $T$
means the past (e.g., $T=-0.27$ means 0.27 Gyr ago).
The locations of the  tips  of the LAs are  marked by ``LA''
for clarity. The dotted line in each panel represents the
Galactic plane (i.e., $l=0^{\circ}$).
Only particles that are within 75 kpc from the Galactic center
are shown so that the evolution of the LAs can be
much more clearly seen (owing to a smaller particle number).
Note that the LAs composed of two streams can pass through
the Galactic plane  around 0.21 Gyr ago.
\label{fig-1}}
\end{figure}

\clearpage
\begin{figure}
\epsscale{.50}
\plotone{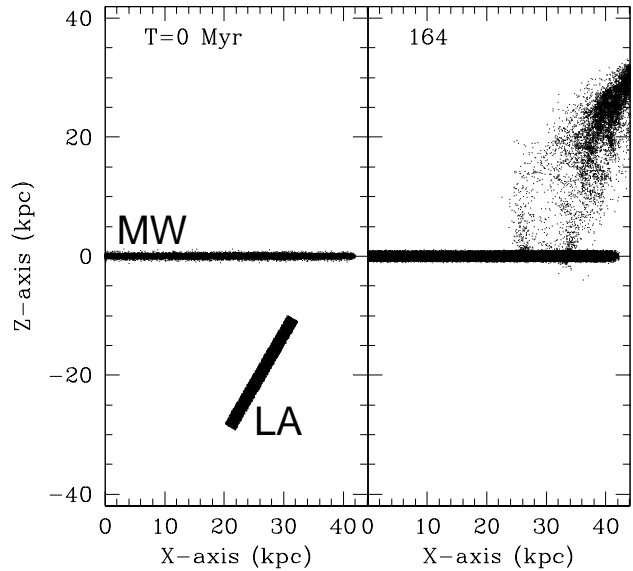} 
\caption{
The initial (left) and final (right) distributions of gas
projected onto the $X$-$Z$ plane
in the Model A for the last $\sim 164$ Myr.
For convenience, the time $T$ in this second set of
simulations  represents
the  time that has elapsed since
the simulation starts (Note that the time definition
is different from that used in Figure 1).
It should be stressed here that the initial 
location (and configuration)
of the LA with respect to the Galactic disk
is chosen such that the physical effects of
the Magellanic impact can be more clearly seen
in this idealized model:
The adopted initial location of the LA
is not exactly the same as that predicted in
the model shown in Figure 1.
\label{fig-4}}
\end{figure}

\clearpage
\begin{figure}
\epsscale{0.5}
\plotone{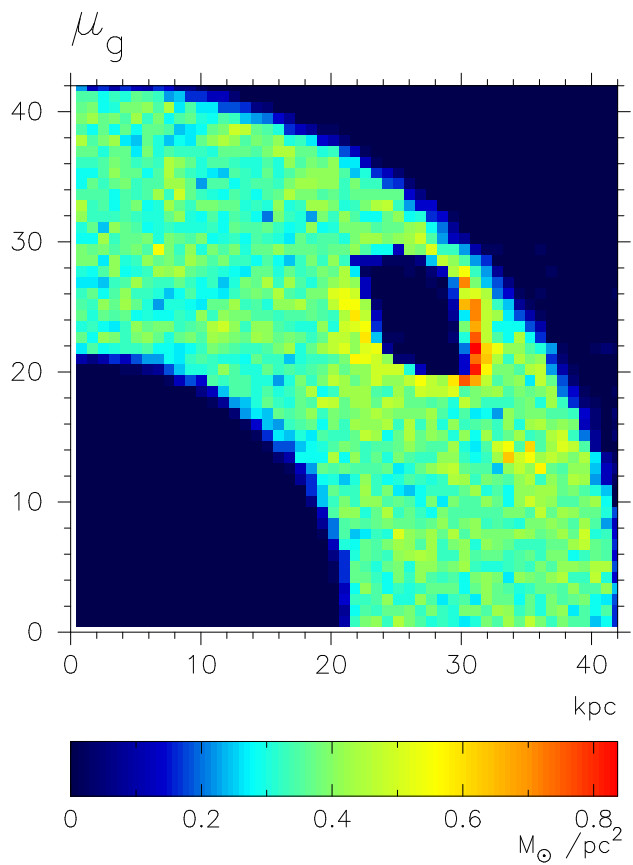} 
\caption{
The two-dimensional
surface gas density distribution (${\mu}_{\rm g}$)
of the outer H~{\sc i}  disk of the Galaxy
projected onto the $X$-$Y$ plane in the Model A 
at $T=197$ Myr (corresponding roughly to the present distribution
of the Galactic H~{\sc i}  gas after the Magellanic impact $\sim 0.2$ Gyr ago).
Only gaseous particles initially in the H~{\sc i}  disk of the Galaxy
are shown so that ${\mu}_{\rm g}$  of the disk can be more clearly
seen. A giant H~{\sc i}  hole with a  high-density ridge in the right side
of the hole can be clearly seen.
\label{fig-5}}
\end{figure}

\clearpage
\begin{figure}
\epsscale{0.5}
\plotone{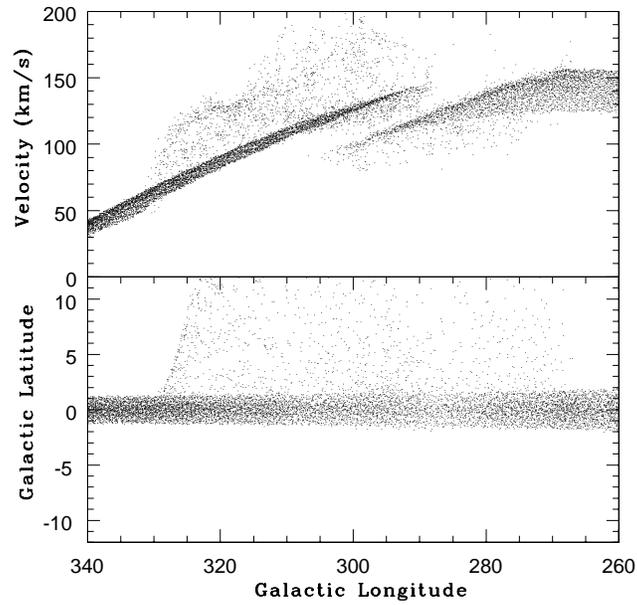} 
\caption{
The simulated H~{\sc i}  longitude-velocity $(l-v)$ diagram (upper)
and the distribution of gas particles on the $(l-b)$ plane (lower)
for the Model B in the present study.
Only gas particles located in the outer Galactic gas disk
(19 kpc $ \le R \le$  28 kpc) are selected so that
the results can be more consistent  with observations
shown in Figures 1 and 2 of M04.
\label{fig-5}}
\end{figure}
\end{document}